\documentclass[11pt,a4paper]{article}
\usepackage{jheppub}

\newcommand{\cO}{{\cal O}}

\newcommand{\nn}{\nonumber}
\newcommand{\vphi}{\varphi}
\newcommand{\Las}{\Lambda^2}
\newcommand{\ve}{\varepsilon}
\newcommand{\IM}{\mbox{\rm Im}}
\newcommand{\RE}{\mbox{\rm Re}}

\newcommand{\eqn}[1]{(\ref{#1})}
\newcommand{\mev}{\mbox{\rm MeV}}
\newcommand{\gev}{\mbox{\rm GeV}}

\newcommand{\smvs}{\vbox{\vskip 8mm}}

\newcommand{\aGG}{\langle\alpha_s GG\rangle}

\newcommand{\gsim}{~{}_{\textstyle\sim}^{\textstyle >}~}

\subheader{UAB-FT-688}

\title{What two models may teach us about duality violations in QCD}

\author{Matthias Jamin}

\affiliation{Instituci\'o Catalana de Recerca i Estudis Avan\c{c}ats (ICREA),\\
             IFAE, Theoretical Physics Group, UAB,
             E-08193 Bellaterra, Barcelona, Spain}

\emailAdd{jamin@ifae.es}

\abstract{Though the operator product expansion is applicable in the
calculation of current correlation functions in the Euclidean region, when
approaching the Minkowskian domain, violations of quark-hadron duality are
expected to occur, due to the presence of bound-state or resonance poles. In
QCD finite-energy sum rules, contour integrals in the complex energy plane
down to the Minkowskian axis have to be performed, and thus the question
arises what the impact of duality violations may be. The structure and possible
relevance of duality violations is investigated on the basis of two models:
the Coulomb system and a model for light-quark correlators which has already
been studied previously. As might yet be naively expected, duality violations
are in some sense ``maximal'' for zero-width bound states and they become
weaker for broader resonances whose poles lie further away from the physical
axis. Furthermore, to a certain extent, they can be suppressed by choosing
appropriate weight functions in the finite-energy sum rules. A simplified
Ansatz for including effects of duality violations in phenomenological QCD
sum rule analyses is discussed as well.
}%


\begin{document}
\maketitle

\section{Introduction}\label{sect1}

In phenomenological applications of QCD, the operator product expansion (OPE)
\cite{wil69,wit77} plays an essential role. It is for example employed in
applications of QCD sum rules \cite{svz79} or the analysis of the $\tau$
hadronic width \cite{bnp92}. Considering in particular two-point correlation
functions of mesonic currents, strictly speaking the OPE is only appropriate
in the Euclidean domain where $s\equiv q^2$ is sufficiently large and negative,
with $q$ being the four-momentum transfer between the two currents. In the
physical, Minkowskian domain with positive $s$, generally bound states or
resonances are present which are governed by long-distance physics and which
cannot be captured by a short-distance expansion.

In the Euclidean domain, a {\em duality} exists between descriptions of the
system in terms of the quark and gluon degrees of freedom of QCD, or the
physical hadronic degrees of freedom. Upon analytic continuation of correlation
functions towards the Minkowskian region, due to the presence of bound-state
poles, violations of the quark-hadron duality are expected to show up
\cite{shi94,shi95,cdsu96,bsz98}.\footnote{For a review the reader is also
referred to ref.~\cite{shi00}.} Since until now no analytic solution to QCD
has been found -- and it is unlikely that this will happen anytime soon -- the
precise functional form of {\em duality violations} (DVs) remains unknown. This
necessitates to resort to the study of models in order to investigate their
influence in phenomenological analyses of QCD. This is particularly pressing
when employing so-called QCD finite-energy sum rules \cite{ckt78,kpt83},
because here contour integrals in the complex energy plane have to be computed
down to the Minkowskian axis.

Based on the models presented in refs.~\cite{shi94,shi95,bsz98,shi00}, the
influence of duality violations on hadronic decays of the $\tau$ lepton has
been investigated in recent years \cite{cgp05,cgp08,cgp09,gpp10a,gpp10b},
because they may play a role in precision determinations of fundamental QCD
parameters, like the strong coupling $\alpha_s$
\cite{aleph05,bck08,bj08,ddhmz08,my08}. For example it has been observed that
the compatibility of values for the gluon condensate $\aGG$, extracted from
fits to vector and axialvector $\tau$ decay spectra \cite{aleph05,ddhmz08},
is not very satisfactory. This poses the legitimate question if the inclusion
of DVs in a consistent analysis may improve the situation regarding the
compatibility of condensate parameters extracted from different channels,
and in how far this would affect the resulting value of $\alpha_s$.

Though the particular model for vector and axialvector correlation functions
employed in the analyses \cite{cgp05,cgp08} is based on features of large-$N_c$
QCD \cite{tho74} and Regge theory \cite{reg59,col71}, it is not directly rooted
in fundamental QCD. Thus, it appears important to detect other systems which
allow for an independent investigation of DVs, in particular their functional
form as well as their possible impact on phenomenological analyses. A system
that allows for an analytical treatment, and bears close relation to quarkonia
in QCD in the limit of a heavy quark mass, is the Coulomb system. In the
context of QCD moment sum rules, to some extent this example was already
studied in ref.~\cite{vol95,eid02}. Furthermore, in relation with hadronic
decays of $B$ mesons, duality-violating contributions resulting from charmonium
states were also considered in refs.~\cite{bbns09,bbf11}.

In the first half of this article, it will be explored in more detail what can
be learned from the Coulomb system regarding violations of duality. To this
end, the notion of duality will be adopted in a more general context. The point
of view taken in the following will be that a certain expansion is employed
which in general can only be considered to be of an asymptotic nature. In our
case this refers to the OPE, and {\em duality} is supposed to imply that the
asymptotic expansion provides an acceptable representation of the full function.
On the other hand, in some regions of the expansion, formally exponentially
suppressed contributions may become relevant, and those will be identified
with DVs.\footnote{This somewhat broader perspective could also be applied to
the perturbative expansion with the ``duality violations'' being the power
corrections due to the presence of QCD vacuum condensates.}

One finding of the study of the Coulomb system below will be that due to the
presence of zero-width bound-state poles on the real $s$-axis, the effects of
DVs turn out to be particularly strong. The same conclusion has been drawn on
the basis of the `t Hooft model \cite{mp09}.  For this reason, the example of
the Coulomb system will mainly be used to gain further general insights into
the structure of DVs. For a more quantitative analysis, in the second half of
this work recourse will again be taken to a simplified version of the model
employed in refs.~\cite{cgp05,cgp08} in the case of a finite width for the
resonances. The finite width entails that the bound state poles move to the
unphysical region of the complex $s$-plane and consequently, DVs in the
physical region turn out to be less pronounced.

In refs.~\cite{cgp08,cgp09} a simplified Ansatz was advocated in order to
incorporate DV contributions into phenomenological QCD analyses. This Ansatz
is supposed to be admissible if the resonances are sufficiently broad and
thus lie sufficiently far away from the Minkowskian, physical axis. Towards
the end of this article, arguments will be given why a related Ansatz should
also be considered on an equal footing which is applicable in part of the
complex energy plane, and the corresponding Ansatz is provided.

\section{Sum rules for the Coulomb system}\label{sect2}

In the following, sum rules for the quantum mechanical Coulomb system will
be set up, which are analogous to the sum rules studied in QCD \cite{svz79}.
Quantum mechanical sum rules had already been investigated in the early years
of QCD sum rules \cite{vzns80,bb81,nsvz81,pt84}, as analytical examples in
which the techniques applied in QCD sum rule analysis could be studied and
tested. Here, this route shall be followed for the question of duality
violations.

Before going into the particular example of the Coulomb sum rule, let us
set up the framework of quantum mechanical sum rules in more general terms
\cite{pt84}.  Consider a particle of mass $m$ under the influence of a potential
$V(\vec x)$. The Schr\"odinger equation for stationary states takes the form
\begin{equation}
\hat H \,\psi(\vec x) \,\equiv\, \biggl[\,-\,\frac{\Delta}{2m}
+ V(\vec x) \,\biggr] \psi(\vec x) \; = \; E\,\psi(\vec x) \,.
\end{equation}
Let $\psi_\alpha(\vec x)$ be the eigenfunction corresponding to the
eigenvalue $E_\alpha$. The resolvent operator $\hat G(z)$ of $\hat H$ is
defined as
\begin{equation}
\hat G(z) \,\equiv\, \Big[\,\hat H-z\hat 1\,\Big]^{-1} \,,
\end{equation}
where $z$ is an arbitrary complex number. Its matrix elements in the
position representation are given by
\begin{equation}
\label{eq:Gmat}
G(\vec x,\vec y;z) \,\equiv\, \langle\vec x|\hat G(z)|\vec y\rangle \; = \;
\sum\limits_\alpha\; \frac{\psi_\alpha(\vec x)\psi_\alpha^*(\vec y)}
{(E_\alpha-z-i0)} + \int \frac{\rho_{\rm cont}(\vec x,\vec y;E)}
{(E-z-i0)}\,{\rm d}E \,,
\end{equation}
where the sum runs over the discrete spectrum, the integral is taken over
the continuous spectrum and $\rho_{\rm cont}(\vec x,\vec y;E)$ is the spectral
density corresponding to the continuous spectrum.

Let us introduce the function $D(E)$ which plays the analogous role of
a physical correlation function in QCD sum rules, for example the Adler
function \cite{adl74}:
\begin{equation}
D(E) \,\equiv \, \biggl[\,\frac{{\rm d}}{{\rm d}E}\,G(\vec x,\vec y;E)\,
\biggr]_{\vec x=\vec y=0} \,.
\end{equation}
Using eq.~\eqn{eq:Gmat}, $D(E)$ can be expressed as
\begin{equation}
\label{DE}
D(E) \,=\, \sum\limits_\alpha\; \frac{|\psi_\alpha(0)|^2}{(E_\alpha-E-i0)^2} +
\int \frac{\rho_{\rm cont}(E')\,{\rm d}E'}{(E'-E-i0)^2} \,\equiv\,
\int \frac{\rho(E')\,{\rm d}E'}{(E'-E-i0)^2}\,,
\end{equation}
with appropriate limits of integration, and where we furthermore have
introduced the full spectral function
\begin{equation}
\label{rhoE}
\rho(E) \,=\, \rho_{\rm pole}(E) + \rho_{\rm cont}(E) \,=\,
\sum\limits_\alpha\, |\psi_\alpha(0)|^2\,\delta(E-E_\alpha) +
\rho_{\rm cont}(E) \,,
\end{equation}
containing both the discrete as well as the continuous spectrum.

Let us now particularise the general expressions to the Coulomb problem. The
ensuing sum rules turn out to be closely related to heavy-quark sum rules in
QCD. Consider two particles of mass $m$ coupled by a Coulomb potential of the
form
\begin{equation}
V(r) \,\equiv\, -\,\frac{\alpha}{r} \,.
\end{equation}
The Coulomb potential is written with a general coupling $\alpha$, though
one may well regard this as the leading term in the heavy-quark potential
containing the QCD coupling $\alpha_s$. Solving the relevant Schr\"odinger
equation, the radial Green function for the Coulomb potential is found to
be \cite{vol95}\footnote{As a matter of principle, also higher-order QCD
corrections could be included in a systematic way. See e.g. ref.~\cite{my98}.
However, for the present purposes, it suffices to stay at the leading order.}
\begin{equation}
G(r,0;E) \,=\, \frac{mk}{2\pi}\,e^{-kr}\,\Gamma(1-\lambda)\,
U(1-\lambda,2;2kr) \,,
\end{equation}
with
\begin{equation}
\lambda \,\equiv\, \frac{\alpha m}{2k} \qquad \mbox{and} \qquad
k \; \equiv \; \sqrt{-\,m(E+i0)} \,.
\end{equation}
$U(\alpha,\beta;z)$ is the confluent hypergeometric function. $G(r,0;E)$
is singular in the limit $r\rightarrow 0$, but the physical correlation
function $D(E)$ remains finite and is found to be
\begin{equation}
\label{MEcoul}
D(E) \,\equiv\, \frac{{\rm d}}{{\rm d}E}\,G(r,0;E) \Big|_{r=0} \,=\,
\frac{m\lambda}{4\pi\alpha}\Big[\, 1 + 2\lambda+2\lambda^2
\psi'(1-\lambda) \,\Big] \,,
\end{equation}
with $\psi(z)\equiv {\rm d}[\ln\Gamma(z)]/{\rm d}z$ being the logarithmic
derivative of the Gamma function. The function $D(E)$ has poles at
$\lambda_n=n$, $n=1,2,\ldots$ which translate to the bound-state energies
$E_n$ of the Coulomb system with
\begin{equation}
E_n \,=\, -\,\frac{\alpha^2m}{4n^2} \,.
\end{equation}
Expanding $D(E)$ around the position of the bound states, $E=E_n-\ve$, from
the leading $1/\ve^2$ singularity the square of the wave function at the
origin can be extracted,
\begin{equation}
|\psi_n(0)|^2 \,=\, \frac{\alpha^3m^3}{8\pi n^3} \,,
\end{equation}
which of course agrees with well known results in textbook quantum
mechanics \cite{gp90}.

Besides the discrete spectrum for $E<0$, in the case of the Coulomb problem
also a continuous spectrum is present for $E\geq0$. Taking the imaginary part
of eq.~\eqn{eq:Gmat}, one observes that the spectral density $\rho(E)$ is
related to the imaginary part of $G(r,0;E)$, namely
\begin{equation}
\rho(E) \,=\, \frac{1}{\pi}\,\IM\, G(0,0;E+i0) \,.
\end{equation}
Again, also $\IM\,G(r,0;E)$ is finite in the limit $r\rightarrow 0$ since
it is a physical quantity. Calculating $\rho_{cont}(E)$ from eq.~\eqn{MEcoul},
yields
\begin{equation}
\rho_{cont}(E) \,=\, \frac{\alpha m^2}{4\pi\Big[1-e^{-\pi\alpha\sqrt{m/E}}\Big]}
\,, \qquad E \, \geq \, 0 \,,
\end{equation}
which is known as the so-called Sommerfeld factor \cite{som31}. Putting
everything together, a finite-energy sum rule, analogous to the case of
hadronic $\tau$ decays, for the quantum mechanical Coulomb problem
can be written down:
\begin{eqnarray}
\label{coulsr}
R^w(E_0) \,\equiv\, \int\limits_{E_1}^{E_0} w(E)\,\rho(E)\, {\rm d}E
&\,=\,&
\frac{i}{2\pi} \oint\limits_{E_0} w(E)\, G(0,0;E)\, {\rm d}E \nn \\
\smvs
&\,=\,&
\frac{i}{2\pi} \oint\limits_{E_0} W(E)\, D(E)\, {\rm d}E \,,
\end{eqnarray}
which defines the moment $R^w(E_0)$ corresponding to an arbitrary analytic
weight function $w(E)$, and the weight function $W(E)$ is given by
\begin{equation}
\label{WE}
W(E) \,=\, \int\limits_E^{E_0} w(E')\, {\rm d}E' \,.
\end{equation}
The central idea of finite-energy sum rules is to employ a phenomenological
representation of the spectral function $\rho(E)$ on the left-hand side of
eq.~\eqn{coulsr} and a theoretically motivated, like the OPE, for $D(E)$ in
the contour integral on the right-hand side. Equating both sides then allows
to infer information on the physical spectrum from the theoretical description,
or to extract theoretical parameters from the phenomenological information. As
most probably in practical applications the theoretical expansion of $D(E)$ on
the right-hand side is only of an asymptotic nature, the consequences for the
finite-energy sum rule need to be investigated.

\section{Asymptotic expansion and numerical analysis}\label{sect3}

For sufficiently large positive or negative energy $E$, that is $|\lambda|<1$,
the $\psi'$-function appearing in $D(E)$ of eq.~\eqn{MEcoul} has a convergent
expansion \cite{as72}. Therefore, in this energy region no duality violations
are expected to occur. They may, however, arise in energy regions where
$\psi'(z)$ only has an asymptotic expansion, which is the case for
$|z|=|1-\lambda|$ going to infinity, while $|\arg z|<\pi$. This happens if
the energy is close to the continuum threshold $E=0$. In this region, the
asymptotic expansion of $\psi'(z)$ is given by \cite{as72}
\begin{equation}
\label{psipas}
\psi'(z) \,\sim\, \frac{1}{z} + \frac{1}{2z^2} + \sum\limits_{n=1}^\infty
\frac{B_{2n}}{z^{2n+1}} \qquad (z\to\infty\;{\rm in}\;|\arg z|<\pi) \,,
\end{equation}
where $B_{2n}$ are the Bernoulli numbers. Even though strictly speaking the
asymptotic expansion \eqn{psipas} should be valid in the full region
$|\arg z|<\pi$, due to the poles of $\psi(z)$ for $z=0,-1,-2,\ldots$, the
asymptotic expansion becomes very inefficient if $z$ approaches these poles.
Quite generally, in the left-half imaginary $z$-plane, that is
$\RE\,z<0$, the asymptotic expansion can be greatly improved
by applying the so-called {\em reflection relation} \cite{as72}
\begin{eqnarray}
\label{refrel}
\psi'(z) \,&=&\, -\;\psi'(-z) + \frac{1}{z^2} + \frac{\pi^2}{\sin^2(\pi z)}
\nn \\
\smvs
\,&\stackrel{z\to\infty}{\sim}&\, \frac{1}{z} + \frac{1}{2z^2} +
\sum\limits_{n=1}^\infty \frac{B_{2n}}{z^{2n+1}} + \frac{\pi^2}{\sin^2(\pi z)}
\qquad (\,\RE\,z < 0\,) \,.
\end{eqnarray}
A more quantitative account on the improvement achieved by including the
additional term in the expansion of \eqn{refrel} will be given in the second
example of the next section.

As the asymptotic expansions of $\psi'(z)$ and $-\,\psi'(-z)+1/z^2$ are
identical, it is clear that the additional term $\pi^2/\sin^2(\pi z)$ has to
be exponentially suppressed. This can easily be verified by rewriting the
$\sin$-function in terms of exponentials, which yields
\begin{equation}
\label{DVterm}
\frac{\pi^2}{\sin^2(\pi z)} \,=\, -\,4\pi^2 {\rm e}^{\pm 2\pi iz} +
\cO\left({\rm e}^{\pm 4\pi iz}\right) \qquad {\rm for}\;\;
\IM\,z\!\!\phantom{x}^{>}_{<}\; 0 \,.
\end{equation}
On the other hand, close to the poles of $\psi'(z)$ the exponentially
suppressed term provides an essential contribution as it gets enhanced by
the poles contained in $1/\sin^2(\pi z)$ for $\IM\,z=0$. Thus, in the general
terminology employed in the present work, the additional term \eqn{DVterm}
can be considered a duality-violating contribution.\footnote{The appearance
of a DV-like term in eq.~\eqn{refrel} is related to the phenomenon of
Stokes discontinuities \cite{din73,pk01}. In the case of the $\psi$- and
$\psi'$-functions the imaginary axis is a Stokes ray, beyond which, that is
in the imaginary left-half plane, the formally exponentially suppressed
terms combine to yield a potentially non-negligible contribution.}

In the following numerical analysis, the FESR \eqn{coulsr} shall be
investigated in a way such that the asymptotic expansion can be performed in
a most transparent fashion. A convenient choice of the complex integration
contour to this end are paths with constant $|1-\lambda|=|z|\equiv z_0$.
The corresponding energy $E_z(\vphi)$ is then given by
\begin{equation}
\label{Ezphi}
E_z(\vphi) \,=\, -\,\frac{\alpha^2 m}{4(z_0{\rm e}^{i\vphi}-1)^2} \,,
\end{equation}
with the parametrisation $z=z_0\,{\rm e}^{i\vphi}$. Because of the square
in the denominator, in the full range of $0<\vphi<2\pi$, the contour
$E_z(\vphi)$ covers the complex $E$-plane twice. However, only the range
$\vphi_0<\vphi<2\pi-\vphi_0$, where the angle $\vphi_0=\arccos(1/z_0)$, and
which contains the section on which the DV term \eqn{DVterm} should be included
in the asymptotic expansion of $\psi'(z)$, provides the correct solution to
the corresponding transcendental equation. Of particular interest are the
two points
\begin{eqnarray}
\label{Ezzero}
E_0 \,\equiv\, E_z(\vphi_0) &\,=\,&
\phantom{-}\,\frac{\alpha^2 m}{4(z_0^2-1)} \,=\, \frac{-\,E_1}{(z_0^2-1)} \,, \\
\smvs
\label{Ezpi}
E_\pi \,\equiv\, E_z(\pi) &\,=\,&
-\,\frac{\alpha^2 m}{4(z_0+1)^2} \,=\, \frac{E_1}{(z_0+1)^2} \,,
\end{eqnarray}
where $E_z(\vphi)$ is real, and which have also been expressed in terms of
the lowest-lying bound-state energy $E_1$. Eq.~\eqn{Ezpi} shows that
in order for the contour  not to hit a bound-state pole, $z_0$ should not
be integer, and that depending on $z_0$, poles with $n\geq{\rm int}(z_0)+2$,
where ${\rm int}(z_0)$ denotes the integer part of $z_0$, need to be included
on the phenomenological side of the sum rule, as they lie inside of the
integration region.

\begin{figure}[!thb]
\begin{center}
\includegraphics[angle=0, width=14cm]{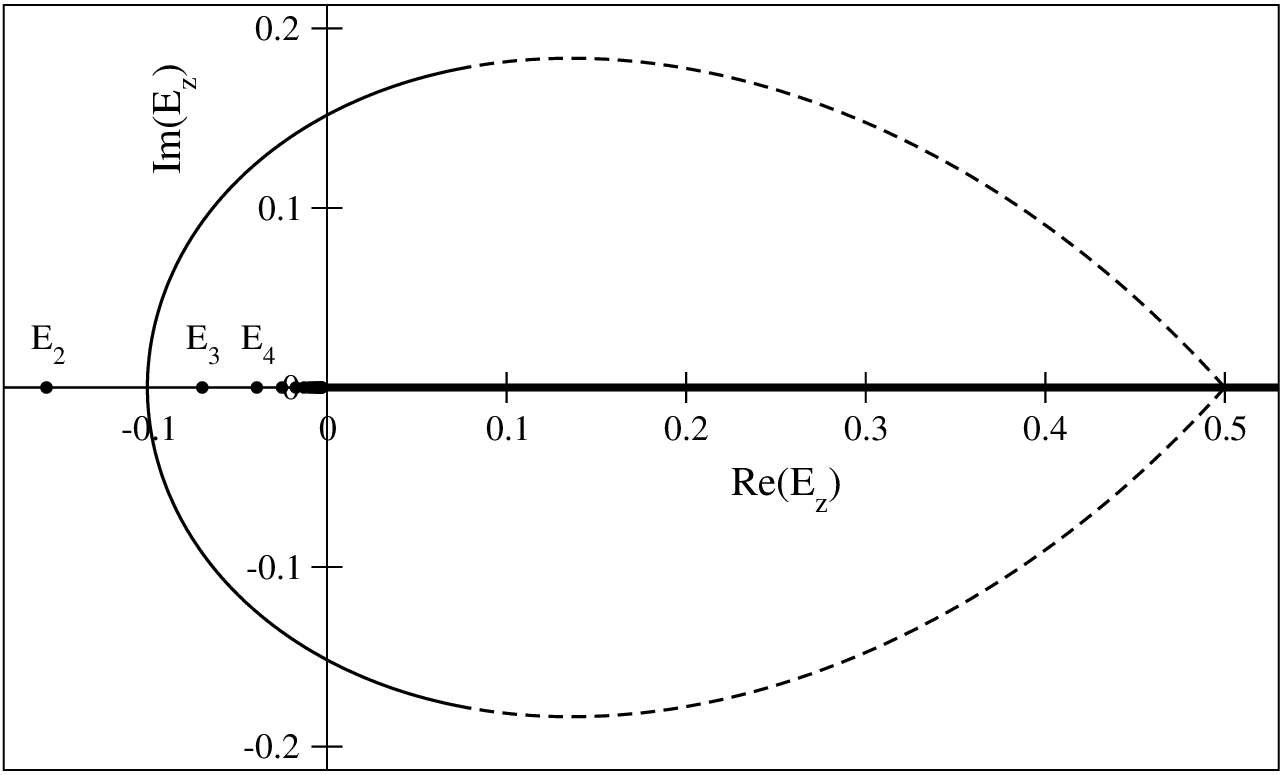}
\end{center}
\vspace{-3mm}
\caption{Integration contour $E_z(\vphi)$ in the complex $E$-plane for the
FESR \eqn{coulsr}, employing $z_0=1.5$ and the further parameters provided in
the text. The part of the contour where the DV term \eqn{DVterm} should be
included has been depicted as the solid line, and the remaining one, where
the asymptotic expansion \eqn{psipas} is sufficient, as the dashed line. Some
bound state poles as well as the continuum cut are also shown as thick dots
and line, respectively.\label{fig1}}
\end{figure}

As a particular example, the sum rule \eqn{coulsr} will now be investigated
numerically for the trivial weight function $w(E)=1$ and the set of parameters
$\alpha=1/2$ and $m=10$. This leads to the lowest ground state being at
$E_1=-5/8$. The integration contour in the $E$-plane is displayed in
figure~\ref{fig1} for $z_0=1.5$, together with some bound state poles as
well as the continuum cut (thick dots and line). The section on which
$\psi'(z)$ admits an asymptotic expansion without DV term is depicted as the
dashed line, while the section where the DV term \eqn{DVterm} should be
included, is displayed as the solid line.

\begin{figure}[!thb]
\begin{center}
\includegraphics[angle=0, width=14cm]{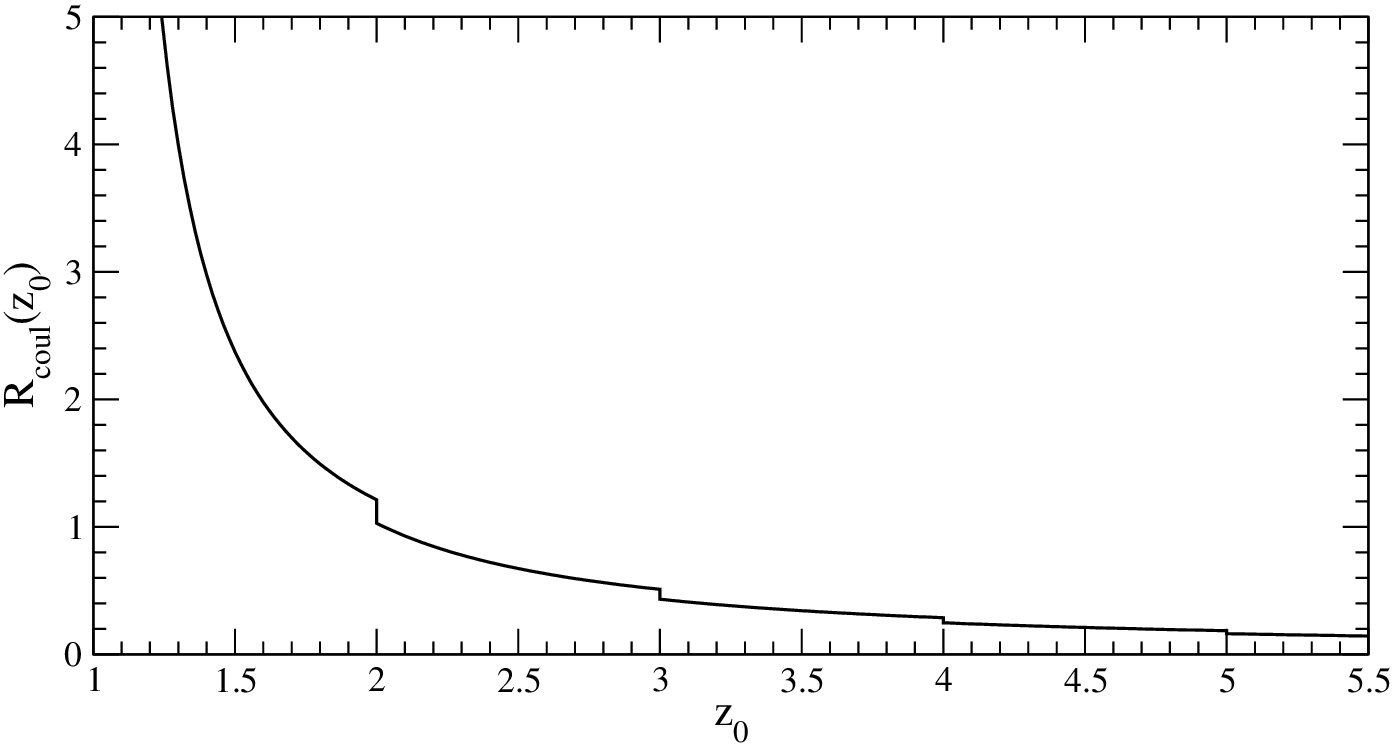}
\end{center}
\vspace{-3mm}
\caption{The solid line represents the moment $R^w(E_0)$ of eq.~\eqn{coulsr}
for the weight function $w(E)=1$, as a function of $z_0$. The discontinuities
appear at the locations where the contour crosses a bound-state pole. The DV
contribution \eqn{DVterm} completely saturates this moment and is visually
indistinguishable from the solid line.\label{fig2}}
\end{figure}

Next, figure~\ref{fig2} shows the moment $R^w(E_0)$ of eq.~\eqn{coulsr} for
the weight function $w(E)=1$, as a function of $z_0$. As expected, at the
locations where $E_z(\pi)$ crosses a bound-state pole, $R^w(E_0)$ is
discontinuous. Employing the full Adler-type function $D(E)$ of
eq.~\eqn{MEcoul}, it is a simple matter to verify that the equality between
the {\em phenomenological} left-hand side of \eqn{coulsr} in terms of $\rho(E)$
and the {\em theoretical} right-hand side is satisfied. On the other hand, when
simulating the OPE with the asymptotic expansions \eqn{psipas} and \eqn{refrel}
it is found that the DV term \eqn{DVterm} completely dominates the moment. At
the lowest considered $z_0=1.2$, the terms polynomial in $1/z$ only contribute
about $0.3\,$\% and at the highest plotted $z_0=5.5$, this contribution is
within the numerical uncertainties of the asymptotic expansion. Hence, in
the respective case, the moment $R^w(E_0)$ is completely saturated by the DV
contribution.

There are two reasons that contribute to this behaviour. On the one hand, the
Coulomb bound states are zero-width resonances and thus the duality violations
are expected to be very strong. In such a case the DVs can be considered
``maximal''. On the other hand, the weight function $w(E)=1$ does not provide
any suppression of the resonance region which could soften the strong duality
violations. The second issue can be investigated further by employing weight
functions which provide some suppression of the resonance region, for example
power-like moments $W^{(k)}(E)=(E-E_\pi)^k (E-E_0)$. Studying these weights
for small $k=1,\,2,\,3$, it is found that indeed at low $z_0$ (close to the
breakdown of the asymptotic expansion) the contribution of the DVs is
suppressed. However, the dominance of DVs again quickly sets in for larger
$z_0$, and even at low $z_0$ the contribution of DVs is still sizeable. The
origin of the latter observation can be traced back to the fact that the DV
term \eqn{DVterm} penetrates some distance into the complex plane, before the
exponential decay becomes effective. Thus, even with the weight functions
$W^{(k)}(E)$, which nullify the contribution on the real axis, the residual
contribution from the full contour integration can remain sizeable.

The dependence of the DVs on the resonance structure could be studied further
by providing the Coulomb bound states with a finite width and varying this
width. However, in view of practical applications of DV models in analyses of
hadronic $\tau$ decays, the following discussion will be continued on the basis
of a second model for DVs, already investigated in
refs.~\cite{shi00,cgp05,cgp08}.

\section{A model for light-quark correlators}

Following refs.~\cite{shi00,cgp05,cgp08}, the structure of DVs shall be
investigated in a second model, which incorporates constraints from Regge
theory \cite{reg59,col71} on the light meson spectrum, and which can be chosen
to roughly resemble the physical spectrum of the light-quark vector current
correlator.  The model employed below is a simplified version of the one
studied in refs.~\cite{cgp05,cgp08}, serving all purposes of the analysis
discussed in the following. Specifically, it is chosen to take the form
\begin{equation}
\label{PiV}
\Pi_V(s) \,=\, -\; \psi\Biggl(\frac{M_V^2+u(s)}{\Las}\Biggr) + {\rm const.} \,,
\end{equation}
where
\begin{equation}
\label{zzeta}
u(s) \,=\, \Las\biggl(\frac{-s}{\Las}\biggr)^{\!\zeta}
\qquad \mbox{and} \qquad  \zeta \,=\, 1 - \frac{a}{\pi N_c} \,.
\end{equation}
As compared to refs.~\cite{cgp05,cgp08}, the lowest lying vector, that is
the would-be $\rho$ meson, is included in the $\psi$-function. The reason
for this will be explained below. Furthermore, in this work the global
normalisation is immaterial and has been dropped. For all remaining parameters,
in the numerical analysis numbers similar to the ones given in ref.~\cite{cgp08}
will be employed:
\begin{equation}
\label{nums}
M_V \,=\, 770\,\mev \,, \qquad
\Lambda \,=\, 1.2\,\gev \,, \qquad
a \,=\, 0.4 \,.
\end{equation}

In order to arrive at an OPE for the model, we require the expansion of the
digamma function at large $|s|$, or correspondingly, large $|u|$. The
expansion in question, which is however only asymptotic, takes the form
\cite{as72}
\begin{equation}
\label{psias}
\psi(z) \,\sim\, \ln z - \frac{1}{2z} - \sum\limits_{n=1}^{\infty}
\frac{B_{2n}}{2n\,z^{2n}}  \qquad (z\to\infty\;{\rm in}\;|\arg z|<\pi) \,.
\end{equation}
In fact, the asymptotic expansion of $\psi'(z)$ of eq.~\eqn{psipas} is of
course just equal to the derivative of eq.~\eqn{psias}. Though in principle
one would have to further expand eq.~\eqn{psias} in terms of powers of $1/s$,
in order to arrive at an OPE-like expansion, this is not necessary for what
shall be discussed in the following, and thus we stick to the asymptotic
expansion in powers of $1/z$.

For finite-width resonances, the poles are on an unphysical sheet, and
$|\arg z|$ never reaches $\pi$. Still, like in the case of $\psi'(z)$,
in the region $\RE\,z<0$ the asymptotic expansion can be substantially
improved by making use of the reflection relation \cite{cgp08,as72}
\begin{eqnarray}
\label{refrel2}
\psi(z) \,&=&\, \psi(-z) - \frac{1}{z} - \pi\,\cot(\pi z) \\
\smvs
\,&\stackrel{z\to\infty}{\sim}&\, \ln z - \frac{1}{2z} -
\sum\limits_{n=1}^{\infty} \frac{B_{2n}}{2n\,z^{2n}}
- \pi\,[\,\cot(\pi z)\pm i\,] \quad
(\,\RE\,z < 0,\, \IM\,z^{\;>}_{\;<}\;0 \,)
\,. \nn
\end{eqnarray}
Again, the logarithm and rational parts of both asymptotic expansions
\eqn{psias} and \eqn{refrel2} agree, and it is a simple matter to convince
oneself that away from the real axis the additional term
$\pi\,[\,\cot(\pi z)\pm i\,]$ is exponentially suppressed, but takes once
more care of the nearby poles.

\begin{figure}[!htb]
\begin{center}
\includegraphics[angle=0, width=14cm]{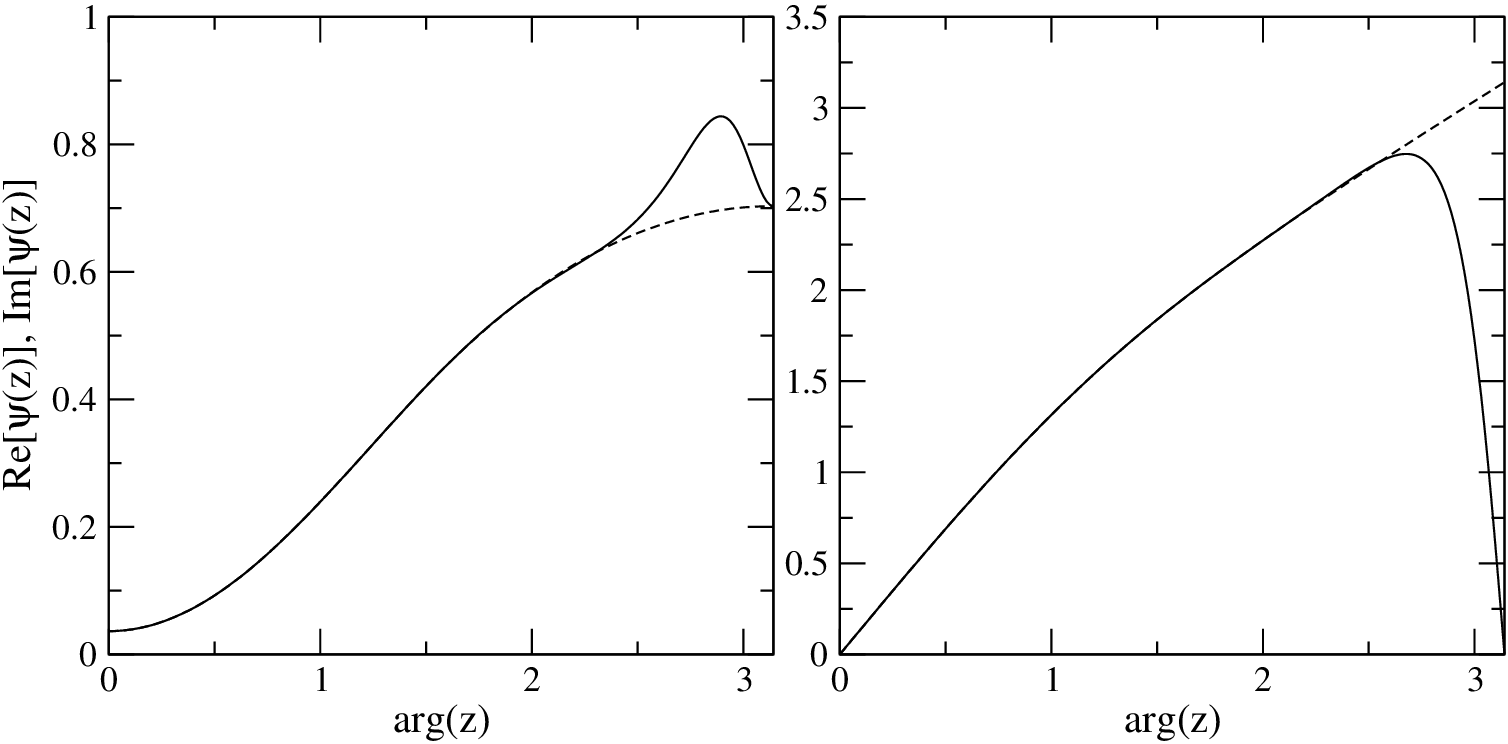}
\end{center}
\vspace{-3mm}
\caption{$\RE[\psi(z)]$ (left pane) and $\IM[\psi(z)]$ (right pane), for
$z=1.5\!\cdot\!\exp(i\vphi)$, as functions of $\vphi=\arg(z)$. The solid
lines correspond to the full result, while the dashed lines show the
asymptotic expansion of eq.~\eqn{psias}, including terms up to $n=5$.
\label{fig2b}}
\end{figure}

This is also demonstrated quantitatively in figure~\ref{fig2b}, where the real
(left pane) and imaginary (right pane) parts of $\psi(z)$ are displayed for
$z=1.5\!\cdot\!\exp(i\vphi)$ as functions of $\vphi=\arg(z)$. The solid lines
correspond to the full function, while the dashed lines show the asymptotic
expansion of eq.~\eqn{psias}, including terms up to $n=5$. For positive real
$z$ at this order the asymptotic series assumes its minimal term. As is
evident, the asymptotic expansion breaks down for $\vphi\gsim 2.5$. The
difference between the total result and the asymptotic expansion is to a very
good approximation covered by the additional term in eq.~\eqn{refrel2}, such
that taking this contribution into account, the quality of the expansion is
analogous to the one for $\vphi<\pi/2$.

To continue, two finite-energy sum rules for the model \eqn{PiV} shall be
investigated. For an arbitrary, analytic weight function $w(s)$, they take
the generic form
\begin{equation}
\label{fesr}
R^w(s_0) \,\equiv\, \int\limits_0^{s_0} \frac{{\rm d}s}{s_0}\,w(s)\,\rho_V(s)
\,=\,
\frac{i}{2\pi}\!\oint\limits_{s_0} \frac{{\rm d}s}{s_0}\,w(s)\,\Pi_V(s) \,,
\end{equation}
where $\rho_V(s)\equiv\IM\,\Pi_V(s+i0)/\pi$, and the contour on the right-hand
side is chosen such that it starts and ends at a real $s_0>0$.

\begin{figure}[!htb]
\begin{center}
\includegraphics[angle=0, width=14cm]{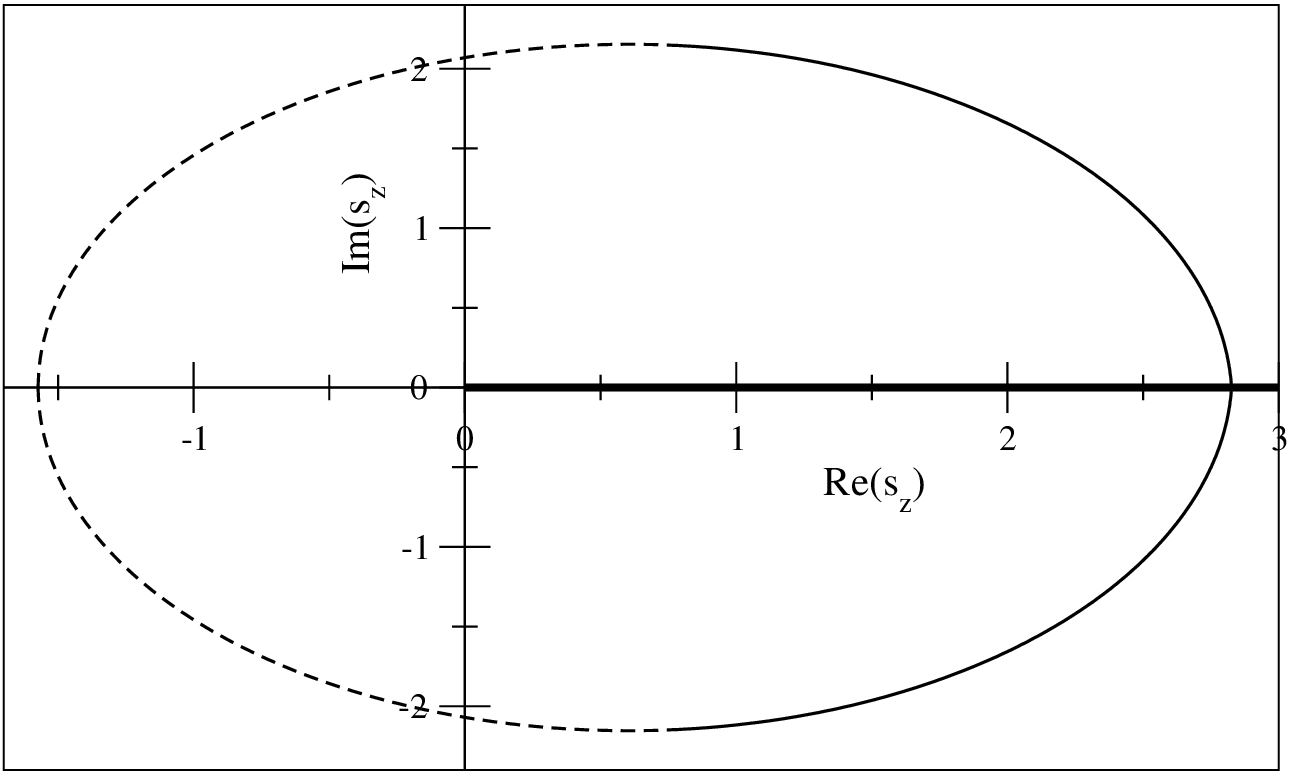}
\end{center}
\vspace{-3mm}
\caption{Integration contour $s_z(\vphi)$ of eq.~\eqn{szphi} in the complex
$s$-plane for the FESR \eqn{fesr}, employing $z_0=1.5$ and the further
parameters provided in the text. The part of the contour where the DV term
should be included has been depicted as the solid line, and the remaining one
as the dashed line. The cut of $\Pi_V(s)$ on the positive real $s$-axis is
also indicated as the thick line.\label{fig3}}
\end{figure}

The asymptotic expansion is again studied most cleanly for contours with
constant absolute value of the argument of the $\psi$-function, $|z|=z_0$.
The corresponding energy variable $s_z(\vphi)$ is then given by
\begin{equation}
\label{szphi}
s_z(\vphi) \,=\, -\,\Las\biggl( z_0\,{\rm  e}^{i\vphi} -
\frac{M_V^2}{\Las} \,\biggr)^{\!1/\zeta} \,,
\end{equation}
and is displayed in figure~\ref{fig3} for the choice $z_0=1.5$. Again, the
portion of the contour on which the DV term should be included is depicted
as the solid line, while the remainder, on which the expansion \eqn{psias}
is admissible, is plotted as the dashed line. In order that the contour
closes below the cut of $\Pi_V(s)$, that is $s_z(0)<0$, one requires
$z_0>M_V^2/\Las\approx 0.41$. On the other hand this entails that
$s_0\gsim 1.2\,\gev^2$.\footnote{This is the reason why the $\rho$-meson
has been included into the $\psi$-function. In the original model of
ref.~\cite{cgp08}, the corresponding requirement would have been
$z_0>M_V^2/\Las\approx 1.6$, which would have necessitated values for $s_0$
at least as large as $s_0\gsim 4.5\,\gev^2$.} Because we have a model with
finite-width resonances and the poles are located on an unphysical sheet,
the angle $\vphi_0$ with $s_z(\vphi_0)=s_0$ is found smaller than $\pi$. For
example in the case of figure~\ref{fig3}, in which $z_0=1.5$, $\vphi_0=2.972$.
In the limit of zero-width resonances, corresponding to the large-$N_c$ limit,
and which in the model \eqn{PiV} is realised as $\zeta\to 1$, $\vphi_0$ would
go to $\pi$ and the poles would lie on the positive real $s$-axis. On the
contrary, for broader resonances which lie further away from the real $s$-axis,
the angle $\vphi_0$ is smaller, and the contribution of DVs gets reduced.

\begin{figure}[!htb]
\begin{center}
\includegraphics[angle=0, width=14cm]{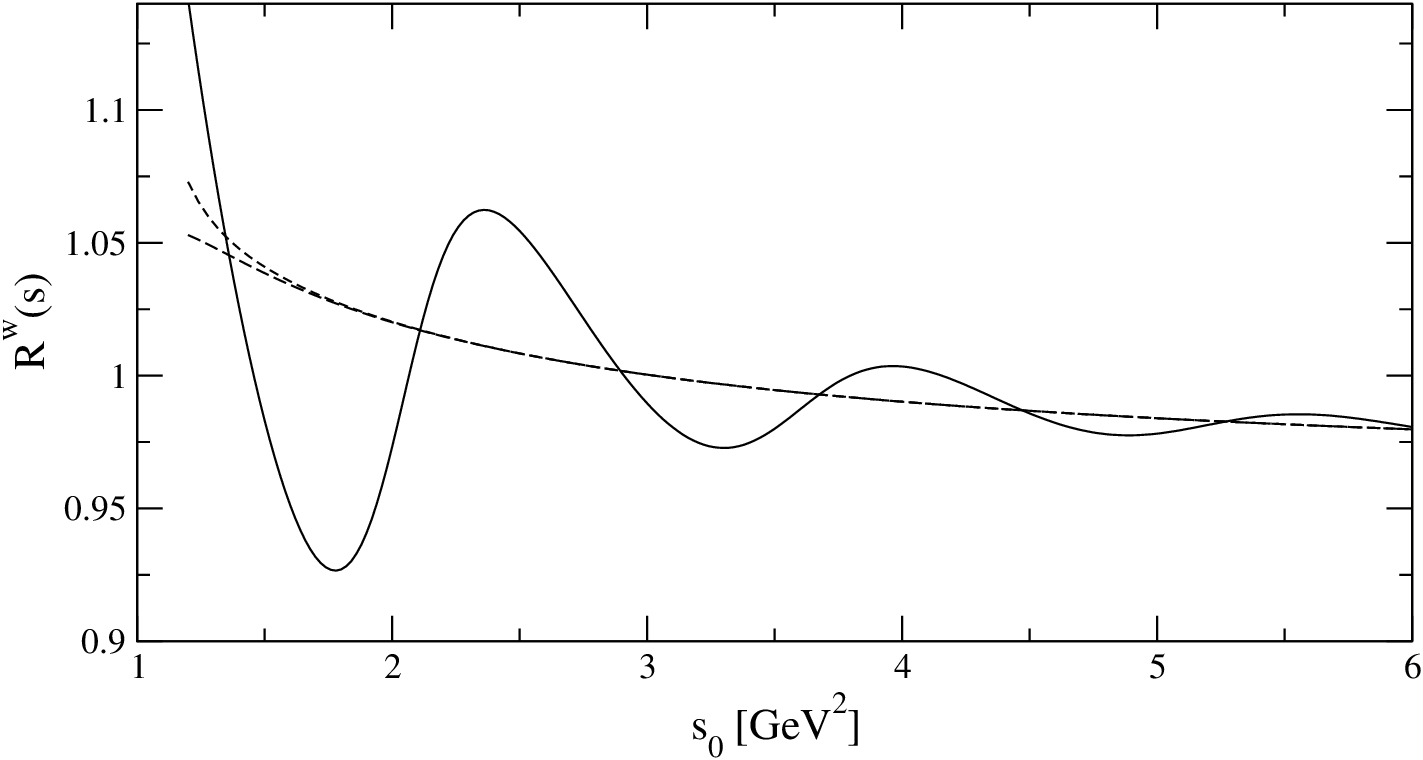}
\end{center}
\vspace{-3mm}
\caption{The solid line shows the moment $R^w(s_0)$ of eq.~\eqn{fesr} for the
weight $w(s)=1$ and plotted as a function of $s_0$. The long- and short-dashed
lines correspond to the power-like asymptotic expansion up to fourth and sixth
order respectively.\label{fig4}}
\end{figure}

The first example of a moment $R^w(s_0)$ corresponding to eq.~\eqn{fesr}
is displayed in figure~\ref{fig4} for the trivial weight $w(s)=1$ and as a
function of $s_0$. The oscillatory behaviour of the moment results from the
presence of resonances, being damped for higher energies. As the long- and
short-dashed lines, the asymptotic expansion \eqn{psias} is plotted up to the
fourth and sixth order in $z$ respectively. As is apparent, the asymptotic
expansion breaks down at about $s_0\approx 1.5\,\gev^2$. Furthermore, large
deviations of the asymptotic expansion to the exact moment are observed, which
are, however, perfectly described by the additional term in eq.~\eqn{refrel2}.
The largest deviation is found at the first minimum around
$s_0\approx 1.76\,\gev^2$, where the DVs amount to about $-11\%$ of the
total contribution.

\begin{figure}[!htb]
\begin{center}
\includegraphics[angle=0, width=14cm]{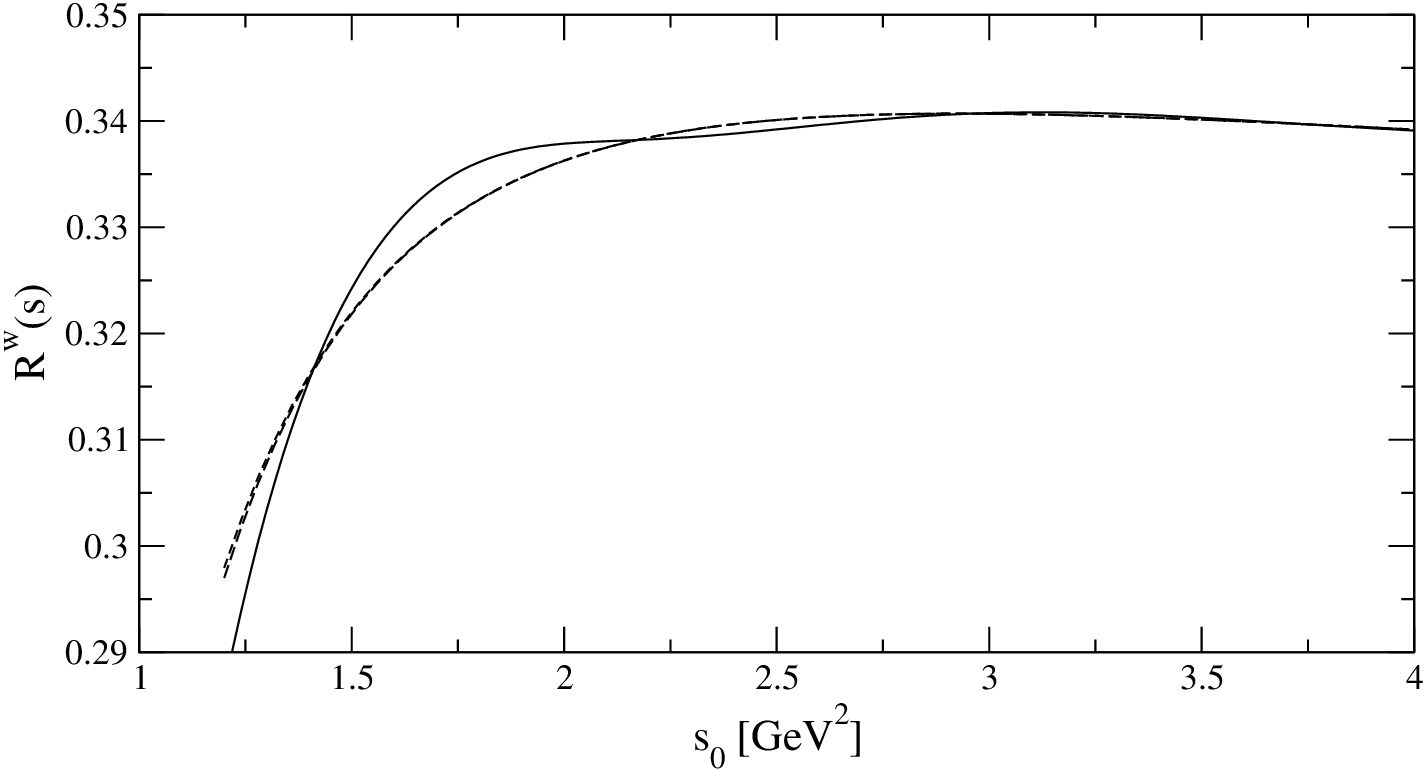}
\end{center}
\vspace{-3mm}
\caption{The solid line shows the moment $R^w(s_0)$ of eq.~\eqn{fesr} for the
weight $w(s)=(1-s/s_0)^2$ and plotted as a function of $s_0$. The long- and
short-dashed lines correspond to the power-like asymptotic expansion up to
fourth and sixth order respectively.\label{fig5}}
\end{figure}

To investigate the influence of weight functions which provide some pinch
suppression at $s=s_0$, in figure~\ref{fig5} the moment $R^w(s_0)$ is displayed
for the weight $w(s)=(1-s/s_0)^2$, which has a double zero at $s=s_0$, just
like the kinematical weight function in hadronic $\tau$ decays. Obviously,
now the influence of DVs is much smaller though still clearly visible. At
$s_0\approx 1.76\,\gev^2$, the contribution of DVs amounts to only $1.6\%$
of the total moment, and a much stronger suppression is observed towards
higher energies.

\section{Conclusions}

In phenomenological analyses of QCD, the operator product expansion is a
widely used tool. While at sufficiently high energies the legitimacy of its
application is unquestionable in the Euclidian domain, the continuation of
the OPE towards the physical, Minkowskian region is inflicted with the
appearance of duality violations. The presence of DV contributions, besides
the OPE, is due to the physical bound states or resonances, whose physics
cannot be captured by the OPE alone.

In the present article, the structure and impact of DVs has been studied on
the basis of two models. Firstly, the quantum mechanical Coulomb system was
investigated, which can serve as a model for heavy quarkonia, and which has
not been considered in the context of finite-energy sum rules and DVs before.
Secondly, a simplified version of a model already discussed extensively in
refs.~\cite{shi00,cgp05,cgp08} was used for comparison with the features
found in the Coulomb model. The latter model incorporates constraints from
large-$N_c$ QCD, as well as Regge theory, and captures the central features
of the light-quark vector or axialvector correlators.

Qualitatively, the physical picture behind the appearance of duality violations
can be summarised as follows: in the Euclidean region, all effects of
resonances or bound-state poles have been sufficiently smoothed out,
quark-hadron duality is expected to work well, and the OPE provides a
satisfactory description of current correlation functions. When analytically
continuing through the complex energy plane towards the physical, Minkowskian
region, the presence of resonance poles becomes more and more prominent,
reflecting itself in a breakdown of the OPE. Mathematically, the DV
contributions show up as formally exponentially suppressed terms in an
asymptotic expansion, which are, however, enhanced by the nearby poles, and
thus can become numerically relevant.

The strength of effects resulting from DVs can be considered ``maximal'' if
the bound states have zero width and the poles lie on the Minkowskian axis.
This case is for example realised in the Coulomb model or when $N_c$ in
eq.~\eqn{zzeta} of the second model is taken to infinity \cite{cgp05}. On the
other hand, the influence of DVs is suppressed for resonances with a finite
width and becomes smaller when the resonances are broader and the poles lie
further away from the physical, Minkowskian axis. This is also seen from the
oscillations in the physical spectrum whose features cannot be described
adequately by the OPE. These oscillations are more prominent for narrower
resonances while they become weaker if the resonances get broader.

If the physical resonances are sufficiently broad, and thus lie sufficiently
far away from the physical axis, a simplified description of DVs as an
exponentially damped oscillatory function appears admissible. Assuming $s$ to
be large enough for the OPE to make sense, motivated by the $\psi$-function
model such a simplified Ansatz was advocated in refs.~\cite{cgp08,cgp09} for
the duality violating spectral function $\rho_{\rm DV}(s)$ in the form
\begin{equation}
\label{rhoDV}
\rho_{\rm DV}(s) \,=\, \kappa\,{\rm e}^{-\gamma s} \sin(\alpha+\beta s) \,,
\end{equation}
where $\kappa$, $\gamma$, $\alpha$ and $\beta$ are a priori free parameters
which in a specific channel have to be extracted from experiment, and which
bear no immediate QCD interpretation. The DV contribution to a moment is then
assumed to be given by
\begin{equation}
\label{RwDV}
R^w_{\rm DV}(s_0) \,=\, -\!\int\limits_{s_0}^\infty \frac{{\rm d}s}{s_0}
\,w(s)\,\rho_{\rm DV}(s) \,.
\end{equation}

In practical applications of QCD finite-energy sum rules, however, the DVs
naturally are required on the complex contour and thus a closely related
Ansatz, which corresponds to the continuation of eq.~\eqn{rhoDV} to the right
complex half-plane, may also be considered:
\begin{equation}
\label{PiDV}
\Pi_{\rm DV}(s) \,=\, \pi\,\kappa\,{\rm e}^{-\,\gamma s\,\pm\,i\,
(\alpha+\beta s)} \qquad
(\,\RE\,s > 0,\,\IM\,s^{\;>}_{\;<}\;0 \,) \,,
\end{equation}
with $\gamma$ and $\beta$ positive. Of course, the $\IM\,\Pi_{\rm DV}(s)$ on
the positive, real $s$-axis reproduces the spectrum of eq.~\eqn{rhoDV}, but
the Ansatz also reflects the form of the exponential suppression of the DV term
given in eq.~\eqn{DVterm}. Both Ans\"atze \eqn{rhoDV} and \eqn{PiDV} would
be totally equivalent, if the DV contribution would be active on all of the
complex contour. However, since the DV term should only be included on part of
the contour (the solid-line sections in figures \ref{fig1} and \ref{fig3}, or,
more generically, on the right-half $s$-plane $\RE\,s>0$), both choices differ
by an exponentially suppressed contribution. As a matter of principle this
should be numerically irrelevant, but it could play a certain role if in
practical applications the exponential suppression is not very pronounced.
This will be studied in the future in investigations like the ones of
refs.~\cite{cgp09,bcgjmop10}.

To conclude, in the presented article observations about duality violations
appearing in asymptotic expansions were summarised on the basis of two models.
One of them has already been discussed in the literature before, while the
other, the Coulomb system, has not been studied with respect to DVs previously.
The qualitative findings of this work should prove useful for phenomenological
QCD studies in the future.

\bigskip
\acknowledgments
I would like to thank Martin~Beneke, Diogo~Boito, Maarten Golterman,
Santi~Peris and Antonio~Pineda for interesting discussions. This work has been
supported in parts by the EU Contract No.~MRTN-CT-2006-035482 (FLAVIAnet),
by CICYT-FEDER FPA2008-01430, as well as by Spanish Consolider-Ingenio 2010
Programme CPAN (CSD2007-00042).

\bigskip

\providecommand{\href}[2]{#2}\begingroup\raggedright\endgroup

\end{document}